\documentclass{aa}
\usepackage{graphicx,epsfig}
\usepackage{natbib}
\usepackage[usenames]{color}
\usepackage{txfonts}
\usepackage{url}
\usepackage[dvips]{hyperref}
\usepackage{stfloats}

\newcommand{\HI}{\ion{H}{i}}
\newcommand{\HeII}{\ion{He}{ii}}

\newcommand{\FeXIV}{\ion{Fe}{xiv}}

\newcommand{\kms}{km s$^{-1}$}

\newcommand{\Rsun}{$R_\odot$}
\newcommand{\YOHKOH}{Yohkoh}
\newcommand{\Hinode}{Hinode}

\begin{document}

\title{EUV jets, type III radio bursts and sunspot waves investigated using SDO/AIA observations}

\author{D.~E. Innes \and R.~H. Cameron \and S.~K. Solanki}
\institute{Max-Planck Institut f\"{u}r Sonnensystemforschung, 37191 Katlenburg-Lindau, Germany}
\offprints{D.E. Innes \email{innes@mps.mpg.de}}

\date{Received ...; accepted ...}

\abstract
{Quasi-periodic plasma jets are often ejected from the Sun into interplanetary
space. The commonly observed signatures are
day-long sequences of type III radio bursts.}
{The aim is to identify the source of quasi-periodic jets observed on
3 Aug 2010 in the Sun's corona and in interplanetary space.}
{Images from the Solar Dynamics Observatory (SDO) at 211\AA\ are used
to identify the solar source of the type III radio bursts seen in
WIND/WAVES dynamic spectra. We analyse a 2.5 hour period during which
six strong bursts are seen. The radio signals are cross-correlated
with emission from extreme ultraviolet (EUV) jets coming from the
western side of a sunspot in AR~11092. The jets are further
cross-correlated with brightening at a small site on the edge of the
sunspot umbra, and the brightening with 3-min sunspot intensity
oscillations.}
{The radio bursts correlate very well with the
EUV jets.
The EUV jet emission also correlates well with brightening at
what looks like their footpoint at the edge of the umbra.
The jet emission lags the radio signals and the footpoint brightening by about 30~s because
the EUV jets take time to develop.
For 10-15 min after strong EUV jets are ejected, the
footpoint brightens at roughly 3~min intervals.
In both the EUV images and the extracted light curves, it looks as though
the brightening is related to the 3-min sunspot oscillations, although the
correlation coefficient is rather low. The only open field near the
jets is rooted in the sunspot.
}
{Active region EUV/X-ray jets and interplanetary electron streams originate
on the edge of the sunspot umbra. They form along a current sheet between the
sunspot open field and closed field connecting to underlying satellite flux.
Sunspot running penumbral waves cause roughly 3-min jet footpoint brightening. The relationship between the waves and jets is less clear.
}


\titlerunning{AR jets and sunspot waves}
\authorrunning{D.~E. Innes \and R.~H. Cameron \and S.~K. Solanki}

\maketitle

\section{Introduction}

X-ray and extreme ultraviolet (EUV) jets in active regions (ARs) are
often associated with repetitive type III radio bursts \citep{CD86,
Kundu95, Chifor08}. Recently they have also been linked
to $^3$He-rich particle events \citep{WPM06, Nitta08}, and to
electron spikes detected at a distance of 1~AU \citep{Klassen11}.

AR jets are mostly seen on the west of the leading spot
\citep{Shimojo96}.
Several were observed by \YOHKOH\ to have
a footpoint on the edge of the sunspot
umbra \citep{Cetal96}.
\Hinode\ has also observed X-ray jets close to the umbral boundary of a sunspot \citep{Nitta08} and
a pore \citep{Chifor08}. Particle acceleration and jet formation is believed to
occur at reconnection sites along current sheets that form between
flux emerging close to the sunspot and adjacent open field
\citep{Setal92, SSH98}, or due to the movement of sunspot satellite flux into
a nearby unipolar flux region with the same polarity as the sunspot
\citep{Cetal96}. In most models the sunspot/pore is part of the jet
environment but plays only a passive role.

In this Letter we show a connection between the sunspot, the
surrounding satellite flux,
 and EUV jets using high cadence (12~s) Solar Dynamics Observatory
 (SDO) images from the Atmospheric Imaging Assembly (AIA) and magnetic field
 observation from the Helioseimic and Magnetic Imager (HMI).
We also investigate (with somewhat inconclusive results) the triggering of jets by sunspot waves.

The chromospheres of sunspots oscillate with a period of 3~min \citep[e.g.][]{Bogdan00}.
 The oscillations were
  first seen as umbral flashes in the chromosphere \citep{BT69}, and are
 believed to be
  magneto-acoustic waves travelling along
magnetic field lines \citep{Centeno06, Bloomfield07}.
In UV transition region lines both intensity and velocity oscillations are observed across entire sunspot umbrae
\citep{Gurman82, Betal99}.
The oscillations also extend into coronal loops anchored in the sunspot
\citep{DeM02b}.

Of special relevance for this Letter is that 3-min oscillations have been observed in microwave flare data \citep{Sych09}, and this was interpreted as flare triggering by 3-min sunspot waves.
Wave-induced reconnection was previously suggested by
\citet{NIS04} when they noticed that in the quiet Sun transition region explosive
events occur in bursts and repeat every 3-5~min. Simulations have shown that the process is viable
\citep{Chen06,Heggland09} but because of the multitude of overlapping small-scale activity along the line-of-sight in the quiet Sun
it is hard to test.
The sunspot-EUV jet relationship discussed here offers a simpler view for assessing the validity of the mechanism.

\section{Observations and data analysis}

\begin{figure}
\centering
\includegraphics[width=\linewidth]{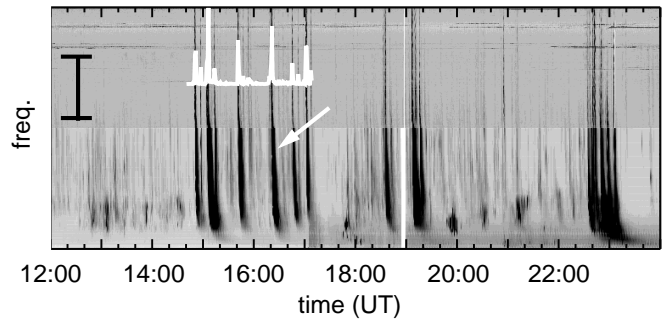}
\caption{WIND/WAVES dynamic spectrum from 12:00-24:00~UT on 3 Aug
2010. The frequency ranges from 20~kHz to 13.6~MHz. The black bar on
the left indicates the frequency range used to obtain the over-plotted
white line which is the signal used for cross-correlation. The white
arrow points to the radio signal from the 16:20~UT jet shown in
Fig.~\ref{onejet_fig}.} \label{radio_fig}
\end{figure}

A series of quasi-periodic type III radio bursts was seen in the
WIND/WAVES dynamic spectrum from 14:00- 24:00~UT on 3~Aug 2010
(Fig.~\ref{radio_fig}). The source of the bursts was jets from the
leading sunspot in AR~11092 (Fig.~\ref{onejet_fig}). This region
produced a C3 flare on 1~Aug 2010, and several other sudden heating
events in the preceding days as described by \citet{Schrijver11}. On
the 3~Aug 2010 the active region was near disk center from the Earth
perspective so SDO had a face-on view of the sunspot.

SDO has two imaging instruments: AIA and HMI. AIA takes images
 of the solar disk through 10 different filters, selected
to single out specific strong lines in the corona and continuum
emission from the lower chromosphere. Photospheric magnetic field,
continuum intensity and Doppler velocity images are obtained by the
optical instrument, HMI.

The analysis presented here uses primarily full resolution AIA
211\AA\ images at a cadence of 12~s and spatial pixel size of
0.6\arcsec. We have also made a movie of the jets by combining a short time series
of 211\AA\
and 304\AA\ with HMI line-of-sight magnetic field images.
The AIA are level 1.0 images, obtained via the Virtual Solar
Observatory and the HMI are level 1.0, obtained from the German
Data Center for SDO. The images have been re-mapped to the same
resolution and de-projected to the start time of the observation
period. The dominant emission in the 211\AA\ images is from \FeXIV\
with a formation temperature $2\times10^6$~K. The 304\AA\ image
selects the transition region, \HeII, line emission with a formation
temperature $5\times10^4$~K.

The visibility of the EUV jets and the sunspot oscillations is
improved in difference images but these are quite confusing when
structures are changing rapidly. We reduce the confusion
by using base difference images. As a
base frame we use the average image over the 2.5~hour period. The
difference is the log of the ratio of the image to the base image.
This brings out changes in faint structures that may be lost in
straight differences.

The radio observations were made with WIND/WAVES \citep{Bougeret95}.
Fig.~\ref{radio_fig} shows the 1~min averaged radio data obtained
from the WAVES archive. The data used for cross correlation is the
signal integrated between 3 and 8~MHz which is indicated by a bar on the
left side of Fig.\ref{radio_fig}. At higher frequencies there is
interference from Earth and at lower frequencies the signal is delayed by more
than a min. Type III radio bursts are caused by electron streams
propagating through the solar
 corona and interplanetary space. The distance from the Sun to the
 radio signal source can be estimated
 from the frequency and a coronal density model. \citet{Leblanc98}
estimate that 13.6~MHz emission is produced at 2~\Rsun. To determine
the electron travel time from the solar surface to 2~\Rsun, we either
have to assume an electron energy or velocity. \citet{Aschwanden02}
gives a typical electron velocity 0.14$c$, where $c$ is the speed of
light for electrons causing 1~MHz emission. Since these electrons
also generate the 13.8~MHz emission, the travel time for
electrons to the site of the observed type IIIs is about 17~s.

\subsection{The origin of the jets}

\begin{figure}
\centering
\includegraphics[width=6cm]{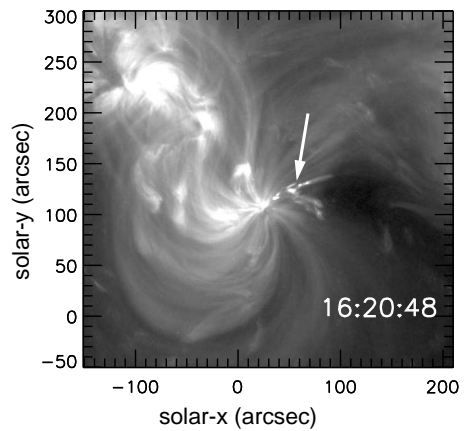}
\caption{ The SDO/AIA 211\AA\ image of the active region and one
of the jets (white arrow). } \label{onejet_fig}
\end{figure}

\begin{figure}
\centering
\includegraphics[width=\linewidth]{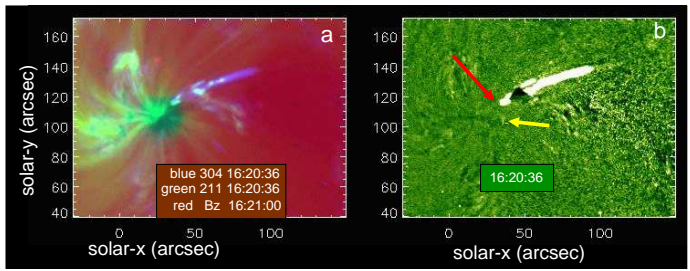}
\caption{(a) A three colour composite SDO image taken from the
accompanying
\href{http://www.mps.mpg.de/data/outgoing/innes/jets/losb_304_211_rd.gif}\tt{movie}.
The EUV 304\AA\ and 211\AA\ filter images fill the blue and green
colours and the red is filled with the line-of-sight magnetic field.
(b) 211\AA\ base difference image. The crest of a sunspot wave is
indicated with a yellow arrow. The red arrow points to a 'footpoint'
brightening.} \label{movie_fig}
\end{figure}

AR~11092 at the time of the
strongest jet is shown in Fig.~\ref{onejet_fig}.
The jets' dynamics can be seen in the
accompanying
\href{http://www.mps.mpg.de/data/outgoing/innes/jets/losb_304_211_rd.gif}\tt{movie}\footnote[1]
{www.mps.mpg.de/data/outgoing/innes/jets/losb\_304\_211\_rd.gif} which
is
 a three colour composite of the AIA 304\AA\
and 211\AA\ filter images and the HMI magnetic field, alongside
the base difference 211\AA\ images. The frame with the strong 16:20~UT jet
is shown in Fig.~\ref{movie_fig}.
All the jets are in the same
direction and appear just north of some short, low-lying loops that
connect the sunspot to a small concentration of satellite flux. This
is similar to the jets described by \citet{Cetal96}. In their model
the triggering of the jets is caused by the outward motion of the
satellite flux driving reconnection at an X-point on the side away
from the sunspot. The jets that \citet{Cetal96} describe were
nearly all accompanied by an H$\alpha$ surge. None of the jets in the
movie have 304\AA\ emission outside the jets which is surprising
because surge material should be visible in this filter. It may be
that surges present in \HI\ are not visible in the 304\AA\ filter or that
surging is common but not fundamental to AR jet dynamics.
Another interesting feature of the AIA jets is bright knots of emission moving along the jets,
simultaneously in all coronal filters. Further analysis is required
to determine whether they are plasmoids formed
at jet onset or due to instabilities in the jet.

In the base difference movie the sunspot waves are obvious over
the sunspot umbra. Close inspection gives the impression that the
waves propagate beyond the umbra along the loops in all directions
probably as slow magnetosonic waves \citep{DeM02b}. The waves
look as though they are propagating across the umbra but the effect
is likely due to wave pulses propagating upwards along the
cone-shaped umbral field which makes the travel time to the observed layer
increase towards the edge of the umbra \citep{Bloomfield07}. Also
conspicuous is recurrent brightening at a small site at the edge
of the umbra in line with the jets. Since this seems to be related
but not actually connected to the jets we call this the footpoint.

\begin{figure}
\centering
\includegraphics[width=\linewidth]{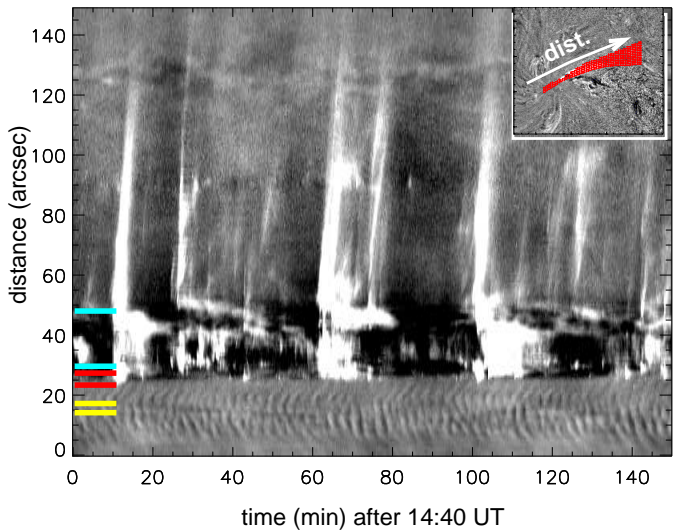}
\caption{Time series of 211\AA\ intensity variations along the
red hashed region shown in the inset. At each distance the intensity
across the region (perpendicular to the arrow) has been averaged. On
the left the yellow and red lines indicate the sunspot and footpoint
regions used for cross-correlation. The light blue lines indicate the
loop region.} \label{jetseries_fig}
\end{figure}

A time series across the sunspot
and along the jet, cutting through the sunspot and footpoint sites
used for the cross-correlation in the next section is shown in Fig.~\ref{jetseries_fig}.
The time series is the average intensity over the width of the
selected region along the cut as indicated. There are $5-6$
strong jets. The strong jets are often followed by a series of weak
jets that fade at shorter distances. Each bunch starts with
brightening in the region bracketed by red bars. A few minutes later the
low-lying loops (light blue region) brighten. The EUV plane-of-sky jet speed ranges from
200-500~\kms\ which is typical of X-ray and EUV jets \citep{Shimojo96}.

\begin{figure}
\centering
\includegraphics[width=\linewidth]{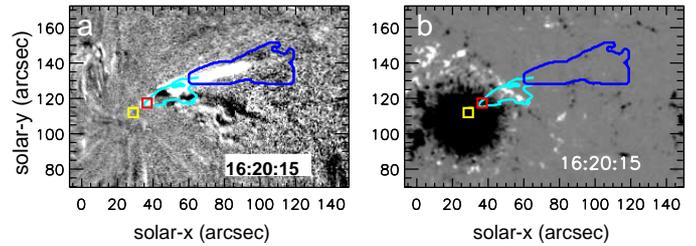}
\caption{(a) Base difference 211\AA\ image (b) Line-of-sight
magnetic field scaled between $\pm250$~G. The
yellow rectangles mark the position used for the sunspot intensity
oscillation, the red rectangles mark the jet footpoint position, the
blue contour outlines the jet region and the light blue the position
of the low-lying loops. These regions are used for the cross
correlation shown in Fig.~\ref{ccjets_fig}.} \label{foursites_fig}
\end{figure}

\subsection{Timing and cross-correlation}
For the analysis here there are three structures of interest (Fig.~\ref{foursites_fig}): the jets
that start just beyond the low-lying loops (blue contour), the
footpoint on the edge of the umbra that produced recurrent
brightenings (red rectangle), and the sunspot umbral oscillations.

It was noted that the footpoint brightens at the time
the EUV jets are observed. An additional question is whether these brightenings and jets
are
excited by sunspot oscillations.
To quantify the
relationship we have cross-correlated the intensity of the footpoint
with the jet and sunspot at the position marked by the yellow rectangle
in Fig.~\ref{foursites_fig}.
The jet intensity is the intensity
integrated within the blue contour.

\begin{figure}
\centering
\includegraphics[width=\linewidth]{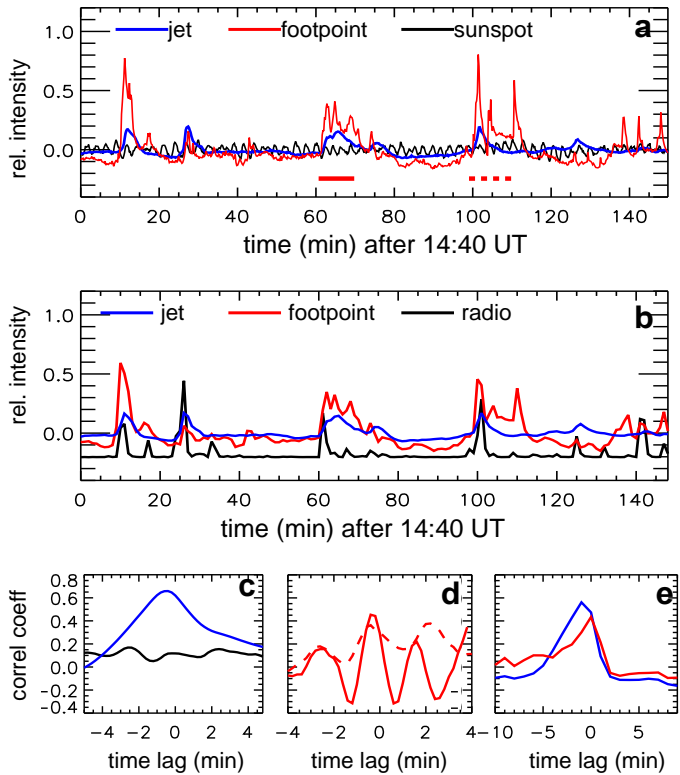}
\caption{Light curves of the sunspot, jet and its footpoint: (a)
211\AA\ light curves with 12~s time cadence of the footpoint,
sunspot, and jet; (b) 211\AA\ jet, footpoint, and radio light curves
averaged over 1~min (c) cross-correlation between footpoint and the
jet (blue) and the footpoint and sunspot (black) over the 2.5~hour
period (d) cross-correlation between the footpoint and sunspot during
two jet intensive periods as indicated by the solid and dashed red
lines in (a). (e) Cross-correlation of the jet (blue) and footpoint
(red) with the radio.} \label{ccjets_fig}
\end{figure}

The temporal evolution of the 211\AA\
 intensity of the different regions is shown in
 Fig.~\ref{ccjets_fig}a. The observations have a time cadence of
 12~s. The 3-min sunspot oscillations (black line) are very
 regular throughout the period. Only the six strong jets (blue line) are visible
 because the weak, diffuse jet signals that are visible in Fig.~\ref{jetseries_fig} are lost when
 integrated over the whole jet region.
 The footpoint variations also suffer from overlapping loop emission and by our choice of a fixed footpoint site. In spite of these drawbacks, there are distinct short period, roughly 3-min, intensity variations at the footpoint
during the strong jet periods.
As one can see from the light curves and confirmed by the correlation coefficient, 0.6, the
 jet-footpoint correlation
 is high. The correlation analysis reveals a 30~s lag of the jet behind the footpoint (Fig.~\ref{ccjets_fig}c).
 Making a cross-correlation over the whole 2.5 hour period, yields
  no correlation between the sunspot waves and the footpoint. But when we focus on two short periods
  with multiple footpoint brightening (one with a
 solid and the other with a dotted red line underneath) the correlation coefficient is 0.4 (Fig.~\ref{ccjets_fig}d).

 In Fig.~\ref{ccjets_fig}b, the intensity variations of the radio
 signal is compared to the footpoint and jet
 intensity. The EUV intensities have been averaged over 1~min
 intervals to match the radio. Both have a high correlation
 coefficient (Fig.~\ref{ccjets_fig}e). The jets seem to lag the radio and the footpoint
 by 1~min, the cadence of the radio data and consistent with the
 30~s time lag between the jets and the footpoint found with the
 higher cadence EUV data. This time lag is
 probably because the jets takes time to develop whereas the other two emissions are compact.
 Fig.~\ref{jetseries_fig} suggests that the jet and the footpoint brightening start together.

\section{Discussion} The SDO images show that the jets causing the
type III radio bursts originate near the sunspot umbra. Investigation
of the source region has shown a strong correlation between the jet
and an EUV brightening at a site on the boundary of the sunspot umbra
suggesting that the jets have a footpoint at the edge of the umbra.
Starting with strong jets, the footpoint intensity oscillates with a
period of about 3 min.
 The footpoint
brightenings are simultaneous with the radio to within 1~min, and precede the
EUV jets' emission by 30~s.

Similar to the X-ray jets studied by \citet{Cetal96} the jets lie
above satellite flux, moving away from the sunspot. Low-lying loops
connecting the sunspot to satellite flux often brighten a few min
after the jet. We suggest these loops
are either newly reconnected loops or loops heated during the
reconnection process.
There are no noticeable surges in the 304\AA\
images so the environment is probably not exactly as observed by
\citet{Cetal96}.

\begin{figure}
\centering
\includegraphics[width=8cm]{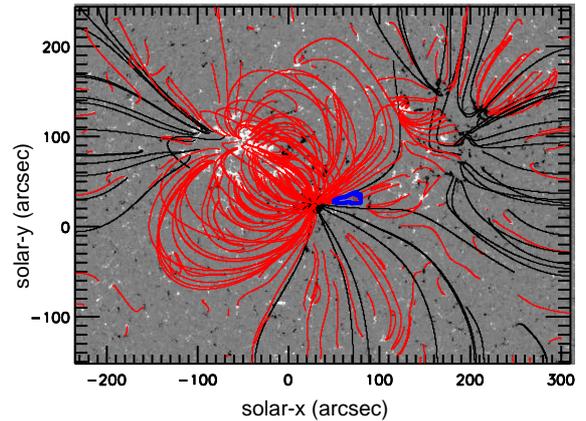}
\caption{Potential field extrapolation of AR~11092.
 Red/black lines are
closed/open field lines projected on the line-of-sight magnetic field. The blue contour outlines the jet region.
} \label{mf_qsl_fig}
\end{figure}

An important difference between these and the \citet{Cetal96} jets
is the magnetic topology. In these jets there is no open field outside the sunspot
beyond the satellite flux. The only open field along which the jets and electron streams can
travel is rooted in the sunspot (Fig.~\ref{mf_qsl_fig}).
 The jets (blue contour) appear between the open (black) and
closed (red) field regions. There are also no signs of flux
emergence in the previous 24~hour, so that the emerging flux trigger
\citep{Setal92} is also unlikely.
Current sheets are however formed along quasi-separatrix layers
between the sunspot open and closed field
\citep{Demoulin96}. Similar changes in connectivity were found at the
base of outflows on the edge of active regions \citep{Baker09}. Given
the right driving forces, reconnection can be triggered along the
current sheet. The fact that we see modulation of the jet footpoint emission
with a roughly 3-min period suggests that sunspot
oscillations may play a role. On the other hand, the enhanced footpoint emission may be due to the superposition of sunspot wave crests on the reconnection emission so at this stage no firm conclusion can be drawn.

With the AIA/SDO images, we are able to observe jets and their
source regions with a temporal and spatial resolution and coverage that allows us
to see connections that had been hitherto hidden. In this letter, we have shown that EUV jets have a footpoint on the edge of the sunspot umbra and propose that sunspot coronal currents create the conditions for reconnection and jets.

\begin{acknowledgements}
We thank Raphael Attie for the interface to the potential field extrapolation code, and the SDO and WIND/WAVES teams for the data. This work has benefitted from discussions with members
of the ISSI, Bern group "Mining SDO data in Europe".
\end{acknowledgements}

\bibliographystyle{aa}


\end{document}